\documentclass[10pt, twocolumn, journal]{IEEEtran}
\usepackage{enumerate}
\usepackage{graphicx}
\usepackage{epsf}
\usepackage{subfigure}
\usepackage{amsmath}
\usepackage{amssymb}
\usepackage{array}
\usepackage{setspace}
\usepackage[amsmath,thmmarks,amsthm]{ntheorem}
\usepackage[ruled,lined]{algorithm2e}
\usepackage{algorithmic}
\usepackage{color}
\usepackage{amsfonts}
\usepackage{dsfont}
\usepackage{xfrac}
\usepackage{breqn}
\usepackage{bbm}
\usepackage{cite}
\usepackage{multirow}
\usepackage{array}
\usepackage{diagbox}

\usepackage{caption}

\graphicspath{{Fig/},{fig/}}

\theoremseparator{.}

\begin{document}

\title{\Huge AI-Enabled Unmanned Vehicle-Assisted Reconfigurable Intelligent Surfaces: Deployment, Prototyping, Experiments, and Opportunities}

\author{\IEEEauthorblockN{\normalsize Li-Hsiang Shen, Kai-Ten Feng, Ta-Sung Lee, Yuan-Chun Lin, Shih-Cheng Lin, Chia-Chan Chang and Sheng-Fuh Chang}}

\maketitle

\begin{abstract}
	The requirement of wireless data demands is increasingly high as the sixth-generation (6G) technology evolves. Reconfigurable intelligent surface (RIS) is promisingly deemed to be one of 6G techniques for extending service coverage, reducing power consumption, and enhancing spectral efficiency. In this article, we have provided some fundamentals of RIS deployment in theory and hardware perspectives as well as utilization of artificial intelligence (AI) and machine learning. We conducted an intelligent deployment of RIS (i-Dris) prototype, including dual-band auto-guided vehicle (AGV) assisted RISs associated with an mmWave base station (BS) and a receiver. The RISs are deployed on the AGV with configured incident/reflection angles. While, both the mmWave BS and receiver are associated with an edge server monitoring downlink packets for obtaining system throughput. We have designed a federated multi-agent reinforcement learning scheme associated with several AGV-RIS agents and sub-agents per AGV-RIS consisting of the deployment of position, height, orientation and elevation angles. The experimental results presented the stationary measurement in different aspects and scenarios. The i-Dris can reach up to 980 Mbps transmission throughput under a bandwidth of 100 MHz with comparably low complexity as well as rapid deployment, which outperforms the other existing works.  At last, we highlight some opportunities and future issues in leveraging RIS-empowered wireless communication networks.
\end{abstract}


{\let\thefootnote\relax\footnotetext
{Li-Hsiang Shen is with the Department of Communication Engineering, National Central University (NCU), Taoyuan 32001, Taiwan. (email: gp3xu4vu6@gmail.com)}}

{\let\thefootnote\relax\footnotetext
{Kai-Ten Feng and Ta-Sung Lee are with the Department of Electronics and Electrical Engineering, National Yang Ming Chiao Tung University (NYCU), Hsinchu 300093, Taiwan. (email: ktfeng@nycu.edu.tw and tslee@nycu.edu.tw)}}

{\let\thefootnote\relax\footnotetext
{Yuan-Chun Lin, Shih-Cheng Lin, and Chia-Chan Chang are with the Department of Electrical Engineering, National Chung Cheng University (CCU), Chiayi, Taiwan. (email: yolen831226@gmail.com, sclinee@ccu.edu.tw and ccchang@ieee.org)}}

{\let\thefootnote\relax\footnotetext
{Sheng-Fuh Chang is with the Department of Communications Engineering, National Chung Cheng University (CCU), Chiayi, Taiwan (email: ieesfc@gmail.com)}}

\section{Introduction}

	The requirement of next-generation tele-traffic is explosively increasing, as the sixth-generation (6G) technology evolves. Such compellingly high demands rely on millimeter wave (mmWave) technology, which has emerged as a critical component for delivering ultra-high data rate services. Given a normal condition, the mmWave transmission between the base station (BS) and the user typically takes place in a line-of-sight (LoS) manner. Therefore, the BS will perform the beamforming technique to align the user without blockage, achieving an exceptionally high throughput. However, non-line-of-sight (NLoS) paths and blockages can significantly hinder mmWave signals in most cases, causing severe transmission disruptions and resulting in weak signals. Therefore, reconfigurable intelligent surfaces (RISs) \cite{acm} are employed to reckon with these new challenges. RISs are widely considered as one of the promising 6G techniques, aiming for extending the service coverage, reducing power consumption, and enhancing system spectral-energy efficiency. The RIS, also known as an electromagnetic surface (ES), is composed of numerous ultra-thin metamaterial-based elements, which can instantly reflect the signals impinging on the surface without additional signal processing. Moreover, as a benefit of cost-effective RIS, it can be readily employed by reconfiguring its phase shifts on each element to artificially alternate the channel between transmitters and receivers.

	RIS-assisted 6G wireless communications have been widely discussed due to its cost-effective and unsophisticated deployment which can be utilized in various systems \cite{mag_ris}. RISs can compensate the NLoS and blockage effects of mmWave transmissions, respectively with reflection and refraction methodologies. Different from conventional relays, RISs can provide significant higher degree of channel diversity compared to the mmWave multi-input-multi-output (MIMO) BS. This enhanced channel diversity enables instantaneous network resilience and adaptability, providing valuable support for network adjustments. The mmWave-RIS relies on the joint optimization design of BS precoding as active beamforming as well as of RIS phase-shifts as passive beamforming. Benefited by minor changes of the protocol, RISs can also assist the existing communication systems, such as non-orthogonal multiple access, full-duplex, coordinate multi-point transmissions. The majority of existing studies that utilize RISs concentrate on various objectives, such as reducing error rates, minimizing interference, lowering power consumption, and maximizing spectral/energy efficiency and physical layer secrecy rates. Recent works have also adopted stochastic theory to analyze the impact of deployment of RISs for various services. More importantly, practical deployment of RISs is quite complex, which largely affects the mmWave system performance. Different orientations of RISs will lead to misalignment of the mmWave beamforming. When is positioned with its back facing the BS, a RIS will be considered a blockage, leading to a lack of signal penetration.

	Moreover, one open issue of RISs is the channel estimation among the BS, RISs, and users. Due to the passive property of RIS elements, the coupled channels cannot be perfectly estimated. This leads to a dilemma in the multi-RIS scenario. Fortunately, artificial intelligence (AI) and machine learning (ML) are considered as the promising techniques in communication systems, relying on the black box operation associated with the predefined input and targeted output. The AI/ML training module is composed of plenty of neural nodes, relying on their non-linear and non-convex calculation. With more features existing from the input, higher performance using AI/ML can be anticipated. Nevertheless, even without estimating the complex RIS channels, we can still dynamically adjust the RIS policy of deployment and configuration solely according to the metrics, e.g., rate, bit error rate, outage, reference signal received power, or signal-to-interference-plus-noise ratio (SINR). Furthermore, the training overhead of AI/ML highly depends on the network scale and performance requirement as well as the selection of AI/ML technique. With more RISs and more configured meta-elements, supervised learning will become infeasible, since it requires excessive data volume and the corresponding ground-truth labels to attain a well-trained neural network. On the other hand, unsupervised learning with incorrect RIS channel estimation as input will lead to inappropriate clustering even with highly correlated data. Alternatively, reinforcement learning can gradually converge to the feasible deployment with comparably low data amount. However, the search space in reinforcement learning is a potential issue to be addressed.

 	To elaborate a little further, conventional manual measurement is laborious and time-consuming. Originated from industrial applications in warehouse, unmanned vehicle is a promising solution. Upon configuring its trajectory, the auto-guided vehicle (AGV) will initiate its automatic measurement and deployment. The robotic system can be embedded with AI/ML for providing intelligent optimal configuration and deployment of RISs. Against this background, in this article we demonstrate the proposed intelligent deployment of RIS (i-Dris) system, including the AGV-assisted RIS platform associated with a mmWave BS and receiver. The federated multi-agent reinforcement learning is designed to automatically adjust the deployment parameters from the BS's feedbacks. We quantify through experiments the employed AI scheme compared to the other existing baselines. Finally, we address some potential future issues in AI techniques and RIS-aided communication system and networks.

\section{AI/ML-Empowered Automatic RIS Deployment}

\begin{figure*}[!ht]
	\centering
	\includegraphics[width=7in]{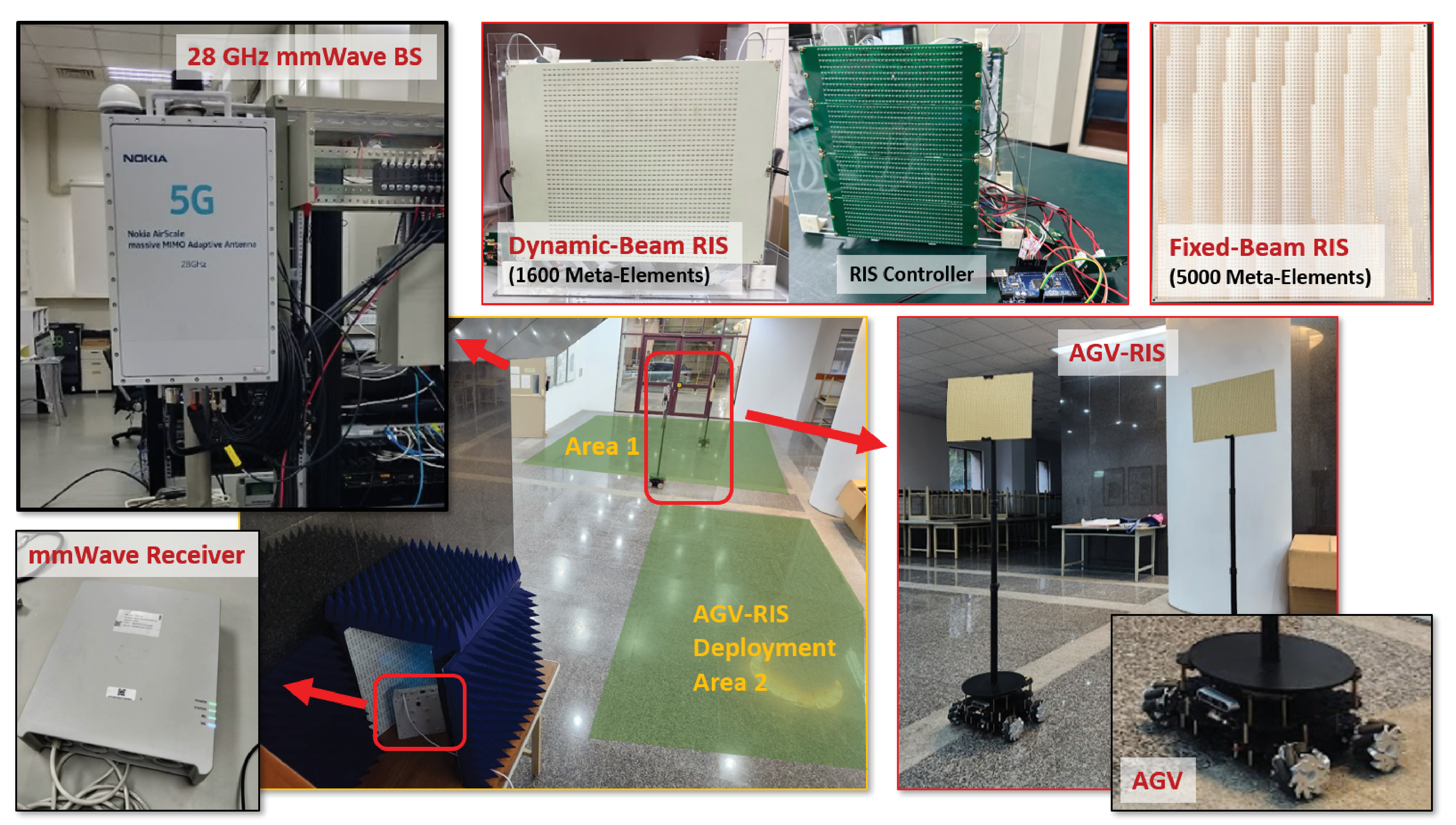}
	\caption{Prototype of the i-Dris system, including a 28 GHz mmWave BS, an mmWave receiver and AGV-RISs associated with their respective deployment area. There show two types of RISs, i.e., dynamic- and fixed-beam based RISs.}
	\label{expscene}
\end{figure*}

\subsection{System Design}

From a system perspective, in Fig. \ref{expscene}, we demonstrate the prototype of the RIS communication system comprising a commercial mmWave BS and RISs. The mmWave BS is capable of performing beamforming to transmit the desired signals received by the mmWave customer premises equipment (CPE) receiver. The fixed-beamforming based RIS is manufactured with material of Rogers 4003, having predefined beam radiation patterns and phase shifts on each meta-material-based element \cite{ccu}. The dynamic-beamforming based RIS provides the $N^{K}$ possible combinations, where $N$ is the controlling bits and $K$ is the number of meta-elements, which possesses comparably higher degree-of-freedom compared to fixed-beam RIS. The RISs are installed on the moving AGV, forming AGV-RIS, whereas the deployment decision is controlled by the AGV. Considering feasibility, we provide a low-complexity setting for AGV movement, with a few discrete actions set in the respective deployment sub-agents of position, height, orientation and elevation: Moving forward/backward/left/right actions are set for position sub-agent; escalation/declination actions are set for height sub-agent; rotating clockwise/counterclockwise are set for orientation sub-agent; increasing/decreasing elevation angles are considered for elevation sub-agents. A device-to-device link using Wi-Fi is conducted between two AGV-RISs, with one of them acting as an anchor. The anchor AGV-RIS is further connected via Wi-Fi to the edge server, whereas the edge is associated with the CPE receiver via fiber links. We further notice an additional edge server will create tunneling (via MobaXterm platform as a remote networking tool) associated with the BS server in order to control the initialization of downlink packet transmissions. We employ \textit{iperf3} built on a client-server model to measure the maximum transmission control protocol (TCP) throughput. Upon initialization, the BS will transmit the mmWave signals to the receiver, with the corresponding throughput logged by the edge server. The edge server will then feedback the performance information to the AGV-RIS for next deployment policy. As shown in Fig. \ref{expscene}, we consider two AGV-RIS deployment areas for addressing the issues of 1st-order and 2nd-order mmWave reflection losses. Upon appropriately deploying RISs in both areas, they can promptly reflect the desired mmWave signal to the user, with the signal loss compromised.

\begin{figure*}[t]
	\centering
	\includegraphics[width=7in]{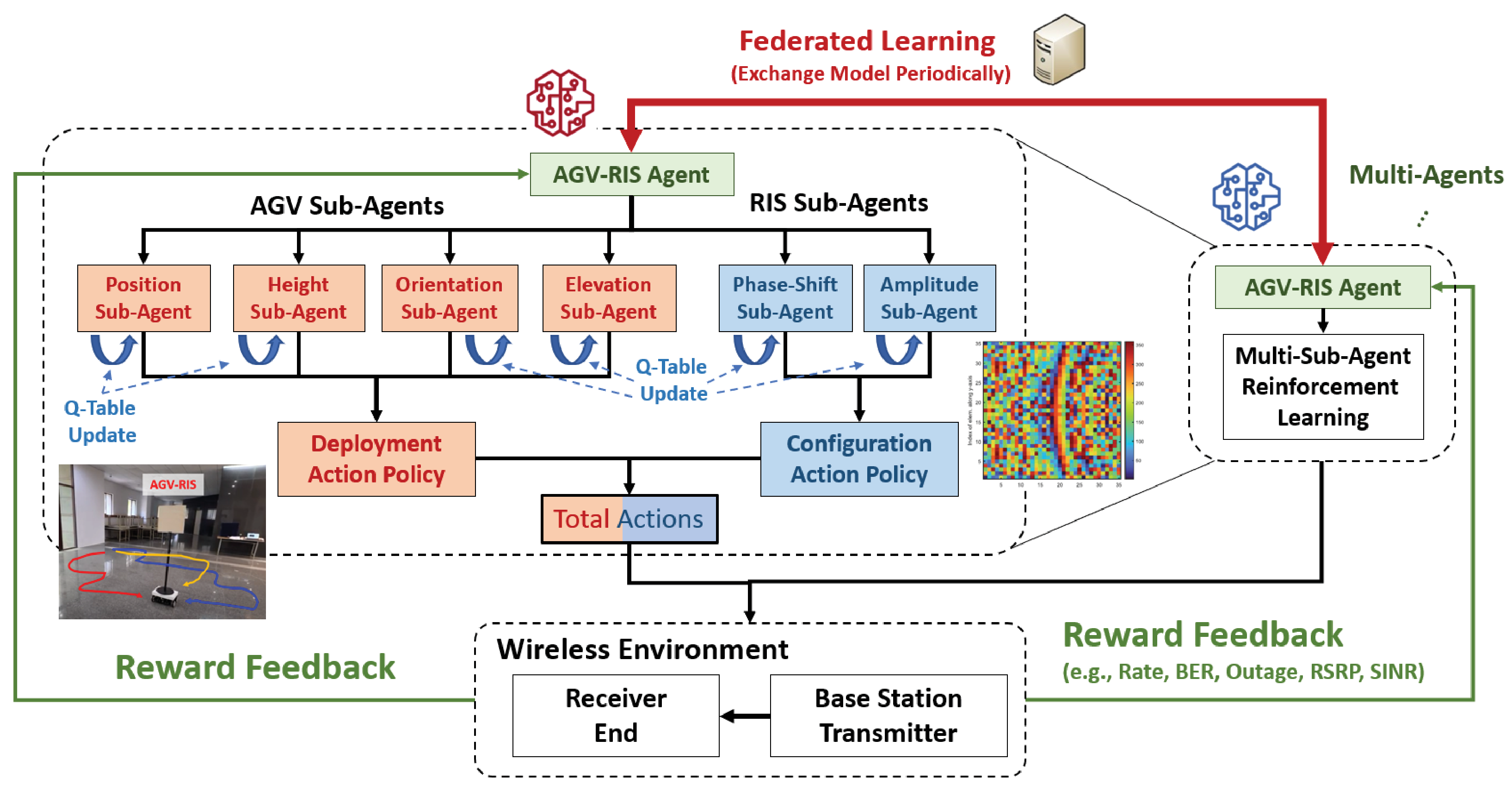}
	\caption{The proposed federated multi-agent reinforcement learning, including multiple AGV-RIS agents. An AGV-RIS consists of AGV and RIS sub-agents, respectively controlling optimal deployment and RIS configuration. The reward associated with the determined joint action is fed back to the individual AGV-RIS for the next decision. Federated learning is performed periodically, having a longer period than reinforcement learning mechanisms.}
	\label{idris}
\end{figure*}

\subsection{Federated Multi-Agent Reinforcement Learning}

From an algorithm perspective depicted in Fig. \ref{idris}, the core of i-Dris is designed with federated multi-agent reinforcement learning (FMARL) for AGV-RIS deployment. Reinforcement learning (RL) constitutes three important factors of action, state, and reward. The action includes AGV-RIS deployment and configuration policies, whereas deployment consists of position/height/orientation/elevation angles of AGVs and configuration has RIS phase-shifts and amplitude coefficients. The subsequent new actions will be exploited upon the observed states and rewards are changed. A greedy-based algorithm is adopted in an exploration-and-exploitation manner. In order to avoid local optimum, a new action will be explored when the randomly-generated number between $[0,1]$ is below the given threshold of the exploration rate, whilst we exploit the action with the highest Q value once the number is above the threshold. However, those actions possess a continuous search space, inducing infinite searching time for obtaining the optimum for multi-AGV-RISs. Therefore, the concept of \textit{multi-agent} \cite{agent} is proposed that hinges on the separate agents associated with the pertinent actions themselves. An individual agent has its own module and table to be updated once the outcomes are unsatisfied. As a result, we have AGV-RIS agent manifesting respective AGV sub-agents of position/height/orientation/elevation angles and RIS sub-agents of phase-shifts and amplitude coefficients, forming a hierarchical multi-agent RL architecture. The joint action will be conducted at AGV-RIS after all actions are collected. Note that the state at each AGV-RIS agent is designed to be its current location and RIS configuration. The reward in RL represents the original communication system performance metric, which can be defined as several factors depending on the requirement, e.g., throughput, bit error rate, outage, received signal strength or signal-to-interference-plus-noise ratio. In our system, we select the throughput as our candidate reward implicitly containing all the other factors. Since deployment has latency owing to signalling overhead, the averaged reward with a given observation window size is considered.

Since all agents have the identical target, i.e., maximizing the throughput, multi-agents perform actions in a collaborative manner termed as multi-agent RL (MARL). Conventionally and efficiently, they just determine their own decision to observe the reward. However, this method potentially leads to the local optimal, since they are guessing others' policies. Therefore, \textit{federated learning} (FL) \cite{fed} can benefit these agents with small amount of information exchanged. The process of FL is that partial module of local RL training is uploaded to the edge, whereas the processed averaged module will be broadcast to all participated agents for next local training. The obtained features contain the other agents' potential information, making it possible to determine a more appropriate joint action. To elaborate a little further, this process will cost additional communication resources. Accordingly, a communication-efficient FL will be employed, with a longer period of model exchange compared to the local RL training time. Benefited by both RL and FL, the proposed FMARL scheme can determine the optimal deployment and configuration action for AGV-RISs. 

During the process, we commence a small number of random actions to collect initial data for AGV-RIS deployment. The agents will interacts with the mmWave BS, RISs, and receiver, allowing us to acquire the throughput reward. The AGV-RIS stores all data in its memory, including current observations and historical deployment outcomes. Based on FMARL, the agents of an AGV-RIS will select their respective action with the highest reward value, whereas the total action will be carried out by the AGV-RIS. Their Q-tables will be updated based on the feedback of throughput reward. After certain time periods, agents will conduct federated learning with parameter averaging to exchange their Q-tables for inferring the hidden information from other AGV-RISs. The procedure of FMARL will be performed until AGV-RISs find a suitable deployment with acceptable throughput performance.

\subsection{Bottlenecks}
There exist several RIS deployment issues, such as multi-network operation, ambient interferences, controls, power issues, data collection and element decays. All these problems potentially affect the throughput and deployment overhead, which are elaborated as follows. 
\begin{itemize}
	\item \textbf{Multi-Network Operation}:
	Our i-Dris system contains fiber links, Wi-Fi links, device-to-device (D2D) links as well as mmWave transmission, forming a compelling complex network. These links potentially incur latency problem owing to communication signalling overhead and computational load when integrating distinct softwares and operating systems. This results in asynchronous connection between BS-RIS-AGV and RIS-AGV-user. Accordingly, the decision is executed by the AGV-RISs with the averaged performance within a windowed duration. Also, the received packets are counted after the completeness of the next deployment of all AGV-RISs. Note that reinitialization of whole setting is required once any of the link failure takes place.

	\item \textbf{Ambient Interference}: 
	As we know, mmWave transmission may be largely affected by objects as well as humans. The walking person will lead to the blocked signal paths, making RISs to re-direct and re-deploy in a feasible place, with more deployment time. Moreover, the weather, humidity and light will induce molecular interference and noise, which leads to signal attenuation losses in both lower-frequency and mmWave bands, causing the longer process or infeasible rates. The feasible range of connection in different layouts should be roughly measured beforehand. To elaborate a little further, the materials of walls will also have impact on the mmWave transmission, e.g., marble surfaces have better reflection coefficient than irregular rocky ones. This will confuse the receiver whether the RISs are accurately deployed and functioning. Hence, absorbers can be strategically deployed to eliminate the undesirable signal paths in real-world scenarios to imitate arbitrary chamber shapes. If the commercial BS is configurable and adjustable, more efficient beamforming algorithm with focused beams becomes essential to address this challenge.

	\item \textbf{Mechanical Control}: 
	The key deployment factors of AGV-RISs hinge on mechanism control of AGVs and circuit processing of RISs. As for AGV, the wheels provide a wide variety of speeds, torque ranges, connector types and mounting options. The conventional fixed straight wheel, same as the tire equipped in a car, can provide an instant action of moving forward and backward. This type of wheels is robust and stable, but limits the deployment of RISs. It requires multiple steps to turn left or right, with a larger mechanical action region. Alternatively, the omnidirectional wheels are capable of instantly moving in all directions in a 360-degree plane \cite{OAGV}. The individual wheel can be configured with different speeds and torque ranges, achieving a compellingly lower latency of deployment. However, lots of mechanism parameters should be adjusted to provide the robust AGV and the corresponding accurate desirable trajectory. Besides the movement, orientation and height are the other controls for AGVs. How to control all these parameters comparable to the signal time scale remains an open challenge. Moreover, the RIS controller is capable of controlling either separate elements or in a patch manner within a second. Although the adopted fixed RIS and one-bit control for RIS have the lower degree-of-freedom for multi-path generation, it can achieve a much lower complexity which is compatible to the wireless signal scale.

	\item \textbf{Power Consumption}: 
The major power issue comes from AGV-RIS, since the moving platform hinges on the battery equipped on the vehicle. The consumed energy includes mechanical behaviors of AGVs as well as the circuit configuration of RISs. It becomes laborious and time-consuming when manually replacing a new battery. Recharging mechanism should be designed. For example, solar panels can be equipped on the AGVs, providing additional efficient energy source. However, it is only applied for the outdoor deployment. Another solution is that AGVs are designed to automatically move to the charging station once below the certain energy threshold. However, it leads to additional expenditure of establishing the charging stations. The other promising solution relies on the wireless power transfer technique. The RISs on AGVs can be configured to absorb partial signals and transform it into energy, namely multi-functional RIS \cite{AGVMF}, while sustaining the moderate rate transmission. Multi-functional RIS possesses the capability of self-sustainability, that is, switching between signal reflection and energy harvesting becomes applicable. Moreover, advanced beam and power allocation schemes can also be leveraged to significantly improve the energy efficiency.

	\item \textbf{Data Recollection}: Owing to the spatial and temporal variation, the performance of RIS deployment may become different. This requires to recollect data or average over the whole dataset. As explained previously, ambient interferences cause spatial variation even with the identical setting. Moreover, data collected from different days potentially becomes invalid due to time variation. Supervised learning will require high data volume to re-adjust its model to mitigate the above effects. However, the reinforcement learning as an alternative solution can adapt to arbitrary environments \cite{DRLRIS}, since training can be performed online.

	\item \textbf{RIS Element Decay}: The elements mostly manufactured in metal on the metasurface may decay from time to time due to oxidation, corrosion or rust, which causes the phase to be fixed or unavailable. It becomes essential to reconfigure the remaining workable phase-shifts to regenerate feasible beamforming pattern. Advanced hybrid materials, such as alloy, compound semiconductor, and nanomaterials can be applied for the surface \cite{NANO}, while sustaining the communication performance of RISs. However, another bottleneck for the fixed RIS is that the decay detection and replacement mechanism are required, since it cannot reconfigure itself.

\end{itemize}

\section{Experimental Results}

\subsection{Experimental Settings}
In Fig. \ref{expscene}, we demonstrate our system prototype, including an mmWave BS, a receiver CPE, RISs (dynamic and fixed beamforming), and AGV-RISs. The commercial BS equipped with functions of central/distributed/radio units is operated at new radio (NR) n257 band at 28 GHz with 4 operating cells. The transmit power is set as 21 dBm, with its peak rate up to around 1 Gbps under a bandwidth of 100 MHz. The mmWave BS supports 32-antenna MIMO capability with the predefined beamforming codebooks associated with at most 32 beam patterns. The beamwidth of each predefined beam is configured around $15$ to $20$ degree. Note that the receiver is also operated at 28 GHz, which is surrounded by the absorbers in order to suppress undesirable signal paths. Two types of RISs are shown at top-right of Fig. \ref{expscene}. The dynamic beamforming of RIS having 1600 meta-elements can control pencil beams with 3-degree beamwidth. Note that the controller will light up the elementwise light-emitting diode (LED) at the back-panel, representing the state of 1, i.e., phase $\pi$. While no-light indicates a 0-state with phase $0$. Moreover, the fixed-beam based RIS with 5000 meta-elements can generate 20-degree beamwidth, which requires no power and controller, with a lighter weight than the dynamic RIS. The two RISs are respectively installed on a pole embedded on top of the omni-wheel based AGVs. Two AGV-RIS deployment areas are with the respective sizes of $12 \times 10 m^2$ and $10 \times 8 m^2$, which represents the first-order and second-order mmWave reflection, with each having one AGV-RIS to move around. The speed of each AGV-RIS is set to 0.3 m/s in all directions. Note that these AGVs will not go beyond its associated area. The distance between the BS to the centroids of area 1 is 10 m and that between the centroids of areas 1 and 2 is 10 m. In the proposed FMARL scheme, we consider two AGV-RIS agents, with each having a greedy-based exploration rate of $0.15$ for preventing local solutions. The period of information exchange for federated learning is set to $5$ action steps in MARL. Discount factor and learning rate are set to $0.5$ and $0.5$, respectively. The historical observation window size is $5$ seconds. Please refer to the video for the proposed i-Dris system in \cite{video}.

\subsection{Results of AI/ML-Empowered Deployment}

\begin{figure}[!t]
	\centering
	\includegraphics[width=3.3in]{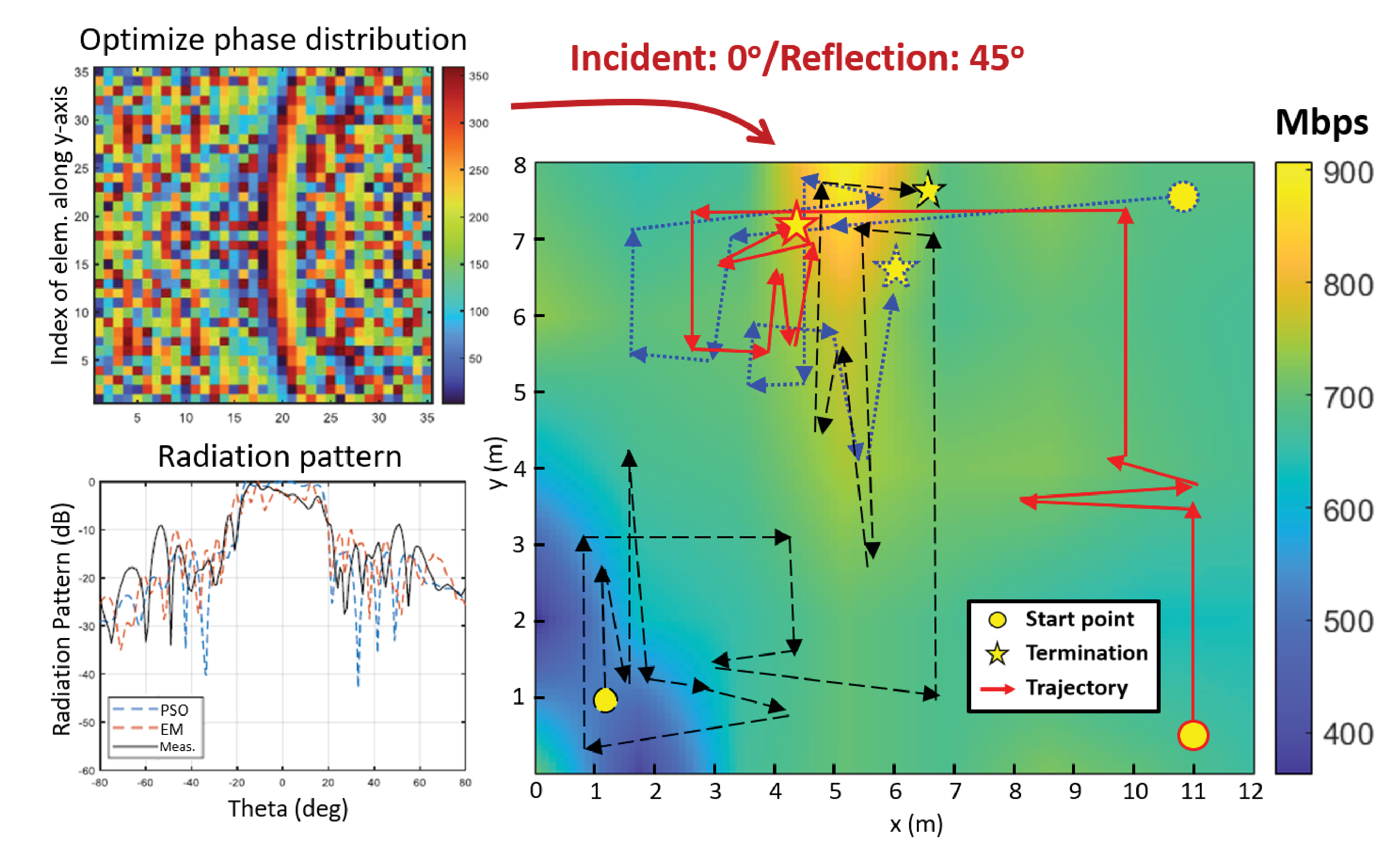}
	\caption{The trajectory of single AGV-RIS experiments for the proposed i-Dris system. Three start points are evaluated with respect to a place near or faraway from the optimal deployment, and initialization with low-rate region. The rate performance of heatmap are offline collected utilizing exhaustive search. Note that the result is conducted, with the phase-shifts of RIS offline determined.}
	\label{exp1}
\end{figure}

\begin{figure}[!t]
	\centering
	\includegraphics[width=3.3in]{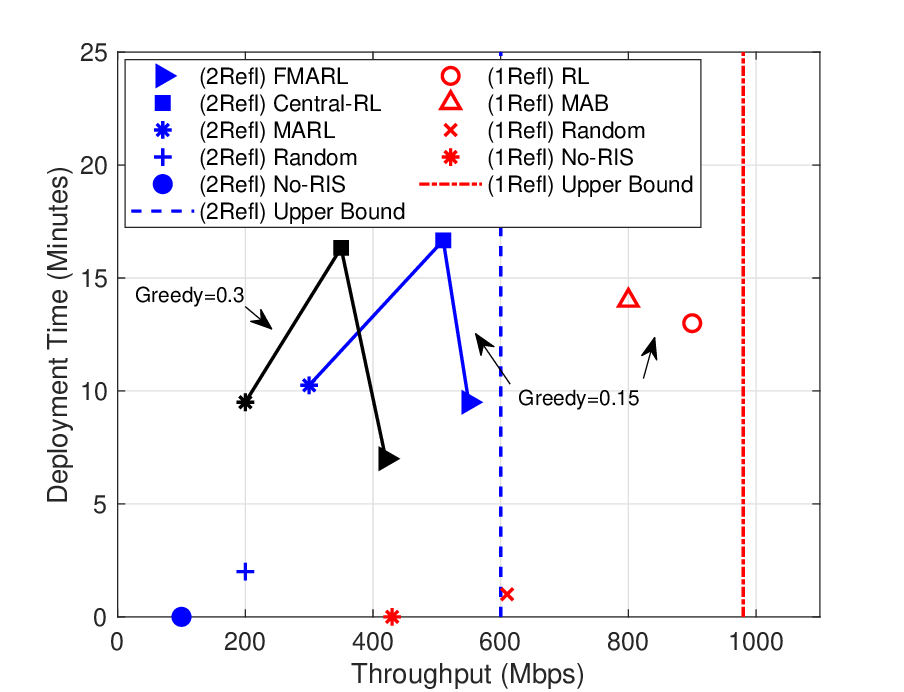}
	\caption{Performance of the proposed i-Dris system compared to the existing benchmarks, including reinforcement learning, multi-armed bandit methods, centralized learning and exhaustive search.}
	\label{exp2}
\end{figure}

In Fig. \ref{exp1}, we conduct three experiments for a single AGV-RIS case, with the depicted simulated pattern and practical dynamic RIS of the optimal phase distribution. We firstly obtain the RIS beamforming patterns with incident angle 0 degree and reflection angle 45 degrees, which is followed by the deployment learning of AGV-RIS. Note that we adopt 1-hour exhaustive search to collect around 100 reference points, with their throughput performance shown as the background heatmap. We can observe that the AGV-RIS gradually converges with the increasing throughput. Our proposed i-Dris can accomplish lower deployment time compared to exhaustive measurement. From the start point with low-rate (bottom-left), i-Dris requires more learning stages to achieve the optimal deployment. As for the moderate start point (bottom-right), AGV moves with a large step size to rapidly search the appropriate trajectory. Therefore, it takes the least time when AGV is initially deployed at a suitable start point (top-right).

In Fig. \ref{exp2}, we have evaluated both deployment experiments of a single AGV-RIS and dual AGV-RISs w.r.t. throughput and deployment time, with the respective upper bounds of rates of 980 Mbps and 600 Mbps. Note that the scenario with 1 mmWave reflection is provided for a single AGV-RIS (only deployment area 1 in Fig. \ref{expscene}), whereas doubled-reflection is for two AGV-RISs (both deployment areas 1 and 2 in Fig. \ref{expscene}). That is, no LoS path exists in the experiments but remains the first-order and second-order reflections. We have compared FMARL in i-Dris to the benchmark schemes of centralized RL, MARL, multi-armed bandit (MAB), conventional RL, random deployment and the deployment without RISs. Note that in centralized RL global actions are  obtained by the edge server. MARL indicates no RL parameter exchange in a multi-AGV-RIS scenario. Different from conventional RL, MAB possesses no state information but only with its reward and the numbers of each action taken. A greedy-based algorithm with exploration probability of 0.15 and 0.3 is compared. It is observed that a higher exploration probability in the greedy algorithm has the worse throughput, since it searches for random action more frequently. For the case of the 1st reflection, we can observe that RL achieves a higher rate and deployment time compared to MAB, since RL possesses additional observation of current and previous states. Albeit little deployment time, the methods of random and no-RIS deployment perform the worst throughput, having the asymptotic result for the 2nd reflection. In the 2nd reflection case, FMARL achieves the highest throughput with the moderate deployment time. While, the centralized RL has lower throughput than FMARL but with compelling deployment time. This is because that the AGV-RISs have to send all information to the server to make the decision, leading to additional high signalling overhead. On the other hand, MARL as the distributed manner without any information exchange requires more iterations to converge, performing the lowest throughput compared to FMARL and centralized RL. It is worth mentioning that conventional convex or metaheuristic optimization techniques are inappropriate in this dynamic problem, because they cannot optimize the instantaneous action owing to the unknown channel information and time-variant performance. Abundant feedbacks with compellingly longer timeslots are required to complete reward calculation of those expired solutions, which becomes not implementable on the proposed i-Dris prototype.

\subsection{Benchmarking}

\begin{figure*}
	\centering
	\includegraphics[width=7in]{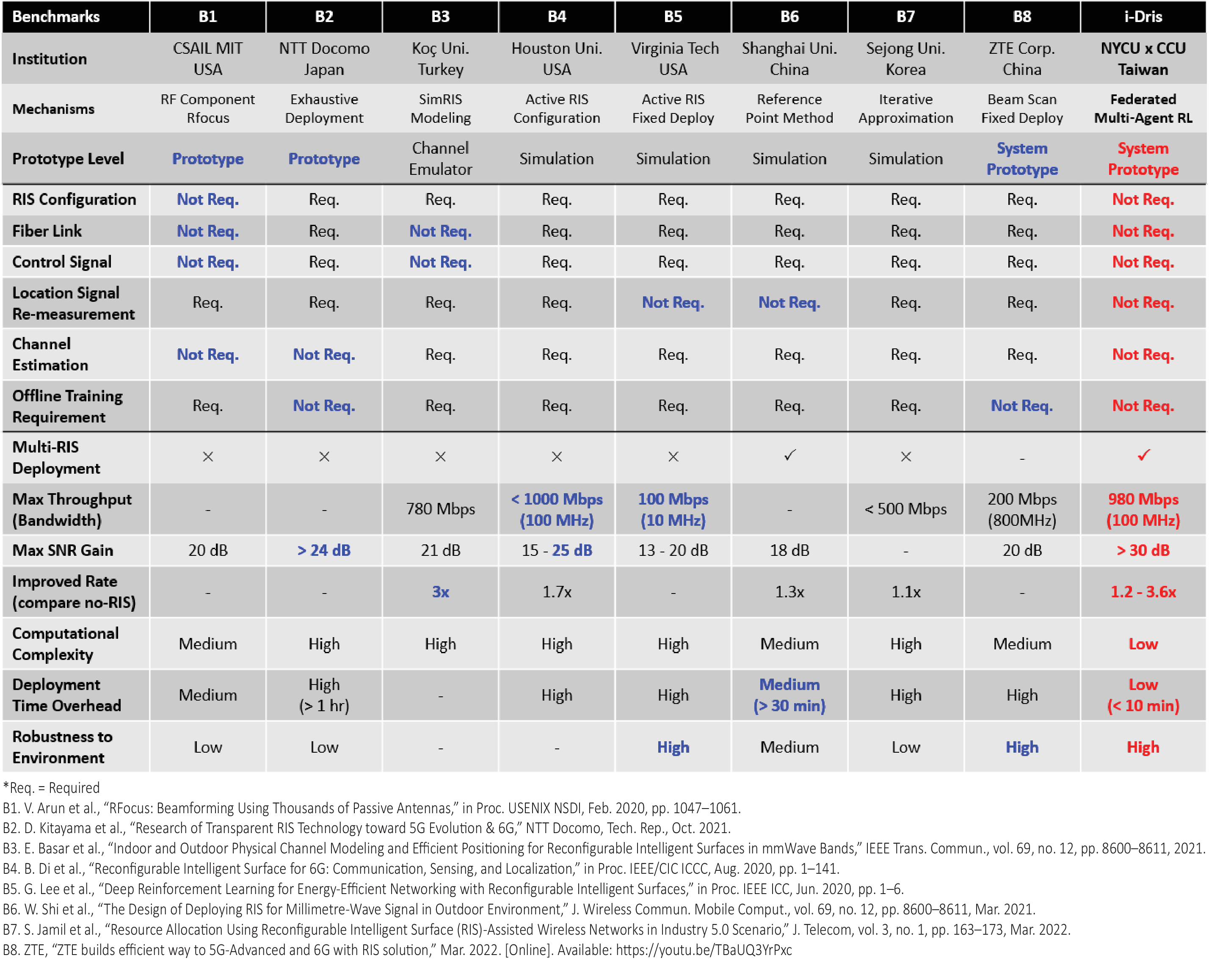}
	\captionof{table}{Comparison of i-Dris with existing works regarding RIS deployment.}
	\label{bm}
\end{figure*}

As shown in Table \ref{bm}, we have provided a table comparison with the existing works regarding the RIS deployment in academia and industry. Most of works focus on either simulation of deployment or implementation of RISs only. In our i-Dris system, a lower overhead can be observed, requiring no control fiber links and signalling, channel estimation and remeasurement as well as time-consuming offline training. i-Dris provides practical multi-RIS deployment, which is not considered in most of current researches. Considering the performance metrics, we have the highest throughput and signal-to-noise ratio (SNR) gain benefited from the joint RIS configuration and deployment. i-Dris achieves 1.2-3.6 times throughput improvement compared to the no-RIS deployment, with low computational complexity and deployment overhead lower than 10 minutes. The proposed i-Dris system can benefit telecom operators, equipment manufactory, and communication industry with economic advantages as follows.
\begin{itemize}
	\item The lower expenditure of infrastructure can be achieved, since we have fewer needs of deploying dense small cells. The network only requires cost-effective RISs to improve signal coverage as well as throughput performance.
	
	\item RIS is an additional infrastructure.
Albeit the high degree-of-freedom for configuration, the dynamic-beam based RISs require additional signalling overhead and protocol adjustment. While, the fixed-beam based RISs are compatible with existing protocols without further amendment.
	
	\item AGV-assisted RIS deployment largely alleviates the requirement of human resources. Upon finishing the setting, i-Dris scheme will automatically search the optimal deployment. Therefore, a large-scale deployment can be conducted with the aid of connected autonomous vehicle.
	
	\item With reinforcement learning, i-Dris capable of supporting rapid plug-and-play deployment without additional high-load computing and laborious learning model training as well as data collection.

	\item Compared to the BS, RIS has a compelling lower power consumption. During the off-peak time, some mmWave BSs can be switched off. Alternatively, we can deploy multi-AGV-RISs to serve users, which conserves electricity with huge cost.

	\item AGV-RIS can be further extended to intelligent services, networks and multiple access in diverse scenarios, such as smart warehouse, private networks, Wi-Fi deployment and radio resource management.

\end{itemize}

\section{RIS-Assisted Wireless Networks}

\begin{figure*}[!t]
	\centering
	\includegraphics[width=7in]{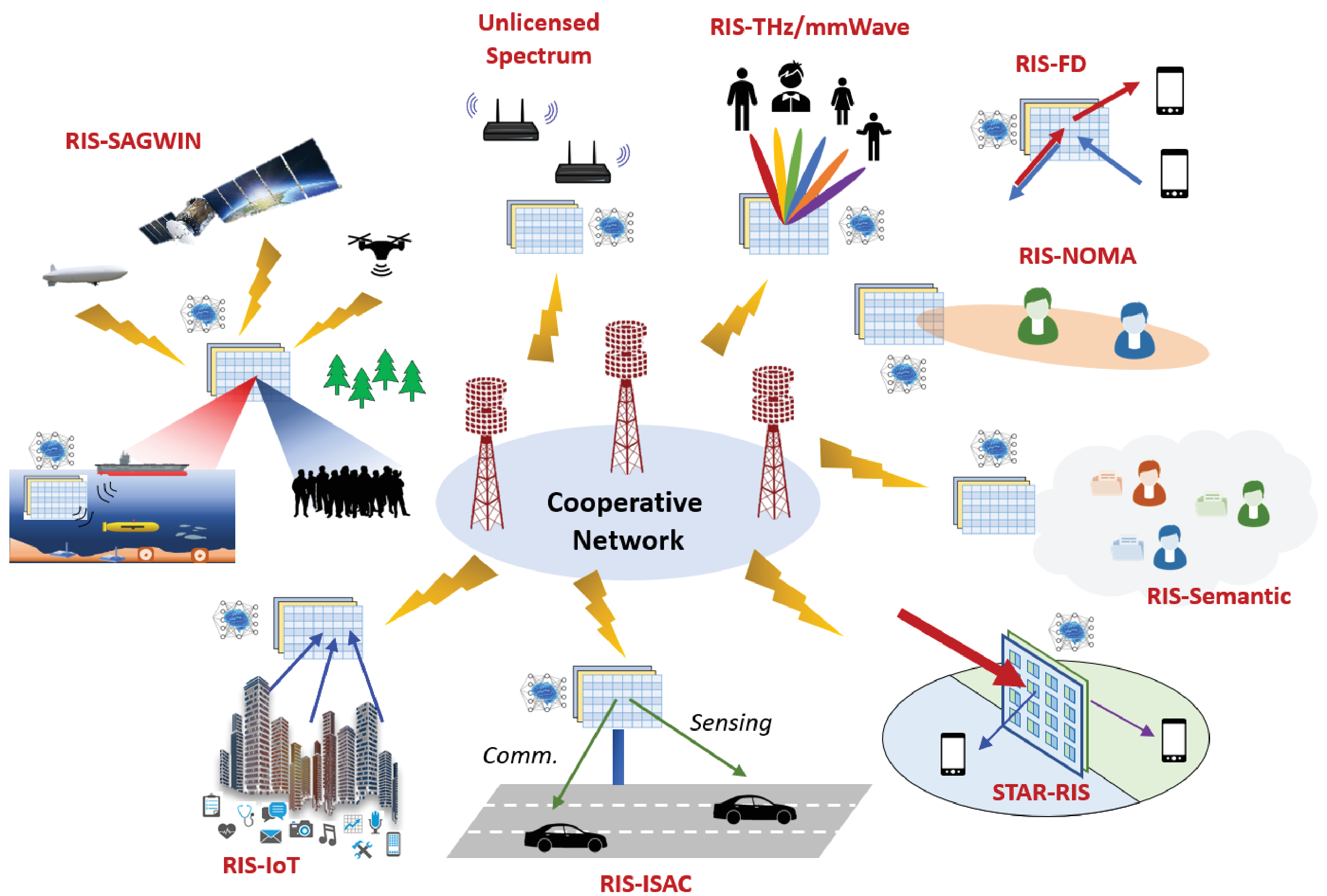}
	\caption{Future cooperative wireless communication networks assisted by RISs. It includes SAGWIN and IoT networks, new ISAC and semantic architectures, STAR-RIS capability, unlicensed spectrum utilization, and new transmission techniques of mmWave, THz, FD and NOMA.}
	\label{future}
\end{figure*}

In Fig. \ref{future}, we have sketched the contours of future cooperative wireless communication networks assisted by RISs. Leveraging RISs in increasingly complex network architectures provokes abundant open challenges. Conventional centralized optimization methods may be inapplicable and unsolvable owing to high computational complexity and highly dynamic environments. Multi-agent system provides parallel low-complexity collaboration, whilst federated RL cooperatively learns the knowledge of uncertain and dynamic environments. Several advanced AI-empowered RIS transmission techniques and emerging new wireless networks are elaborated in detail as follows.

\textbf{Advanced AI/ML}:
Upon having different sizes of multiple RISs for various applications, there may arise complex configurations and deployment for mitigating the excessive interferences. Using conventional spatial-temporal neural network architectures, such as convolutional and recurrent neural networks, it becomes infeasible to address such issue \cite{AIRIS}. Graph attention-based neural networks can take into account the correlation between RISs and between elements, more focusing on the aspects advantageous to the system performance. On the other hand, transformer-based deep neural networks originated from computer vision domain can empower the arbitrary dimensions of inputs and outputs in support of different sizes, types and multi-tasks of RISs. Moreover, hierarchical learning with federated reinforcement learning can also provide a large-scale collaborative RISs-assisted network in both vertical and horizontal services. To perfectly leverage the above-mentioned schemes, several informative features can be hybridly applied, including camera, radar and lidar sensors depending on the affordable expenditure. Accordingly, sensory data fusion can be designed to further enhance the deployment performance.

\textbf{RIS Transmissions}:
RISs can be manufactured to harness multiple frequency bands at the same time, i.e., operating at microwaves, millimeter waves as well as terahertz (THz) \cite{RISTHZ}. It can enhance the signal coverage at lower bands such as indoor Wi-Fi applications for unlicensed spectrum utilization. With more elements to be configured on RISs, it can generate massive pencil-beams in support of massive connected devices, providing ultra-high transmission rate. Moreover, it is capable of supporting different existing transmission techniques, such as non-orthogonal multiple access (NOMA) and full-duplex (FD). In RIS-NOMA, RISs generate multi-paths to assist the signals with different power levels to be well-superposed and transmitted to the respective users, with one strong and the other weak channel. Appropriately deploying RISs can assist a better user pairing in NOMA transmission. In RIS-FD, RISs reassign the potential signal space in order to significantly alleviate the strong self-interference and inter-user interferences for joint uplink and downlink transmissions. The optimal deployment of RISs can provide the highest data rate for both uplink and downlink users while minimizing the RIS-induced self interference. To elaborate a little further, beyond the Shannon capacity theory, RIS-assisted semantic communications can encode data sources into semantic information to transmit. More encoded semantic information can be delivered when the well-deployed RISs provide better channel quality.

\textbf{STAR-RIS Network}:
As an extension of RISs, simultaneously transmitting and reflection RISs (STAR-RIS) \cite{star} allow the impinging signals disseminating in both sides of metasurfaces, providing a 360-degree transmission coverage. This is achieved by configuring dual inductors, capacitors and resistors associated with the doubled-sided lossless materials. The STAR-RIS can be configured in several types of functions, including energy splitting, mode switching, or time-division manner. The STAR-RIS is further extended from a single-sided to a double-sided metasurface, i.e., the signal can arrive in both sides of a STAR-RIS, making it more flexible upon deploying it in any orientation. Furthermore, hybrid RIS and STAR-RIS networks can assist different service demands associated with low traffic loads or hotspots. However, the high degree of freedom of configuration and interference management as well as practical deployment become more complex than that of non-STAR-enabled RIS.

\textbf{RIS-Assisted SAGWIN}:
Space-air-ground-underwater integrated network (SAGWIN) constitutes the global conventional terrestrial and non-terrestrial networks. The high-latency and high signal loss induced by these distant transmissions can be compromised by the deployment of RISs. Low-Earth orbits (LEO) equipped with metasurfaces can collaboratively redirect the impinging signals to the desired distant receiving units on the ground \cite{RISSAGIN}. The deployment of a doubled-RIS architecture with one near the LEO and the other one close to the ground can further enhance the signal strengths at the respective sides. However, it strikes a compelling tradeoff between the performance and the expenditure of massive RIS deployment and maintenance. Moreover, multi-band RISs can be deployed undersea, harnessing the complex underwater channel paths in different depths and frequencies. Benefited by the deployment of multi-hop RISs or STAR-RISs, the signal can be intactly relayed from the space, sky to undersea. Internet-of-Things (IoT) is another important architecture in SAGWIN, which usually contains ultra-massive devices to be associated with the gateway or BS. With the appropriate deployment of RISs, the area with plenty of emergent IoT devices can be well-managed, achieving low latency and low connection failure rate.

\textbf{RIS-Enabled ISAC}:
Integrated sensing and communication (ISAC) is deemed to be 6G potential topics, implemented by dual-function radar communication systems \cite{RISISAC}. Leveraging RISs is capable of generating ample channel paths for detecting multiple objects as well as for transmitting multi-source signals. Partitioning some RIS elements for sensing while the others for transmission become a potential alternative solution. Moreover, benefited by the high-dimensional passive elements, interference amongst the paths induced from radar and communication can be well-managed. The deployment of STAR-RIS with simultaneous reflective and refractive links can provide a full-duplex ISAC with massive objects and joint uplink and downlink users at both sides of metasurfaces.

\section{Conclusions}
	In this article, we have firstly described the benefits of utilizing RISs as well as the employment of AI/ML in RIS configuration and deployment. We have implemented a prototype of i-Dris system which consists of AGV-assisted RISs associated with the commercial mmWave BS and receiver CPE. The i-Dris solution is driven by the federated multi-agent reinforcement learning, intelligently and automatically deploying the RISs according to the rewards given from the BS. The experimental results evaluated the stationary measurement in different factors and scenarios, which is followed by the intelligent deployment of AGV-RIS. The proposed i-Dris system has the advantages of low-cost, low-complexity and rapid-deployment, outperforming the other existing works. At last, we also discussed some future issues when leveraging advanced AI/ML techniques and RISs embracing next-generation networks.

%

\bibliographystyle{IEEEtran}
\bibliography{IEEEabrv}

\end{document}